\documentclass[aps,prd,twocolumn,showpacs,amsmath,amssymb,nolongbibliography]{revtex4-2}
\usepackage{amsmath}
\usepackage{graphicx}
\usepackage{subfigure}
\usepackage{siunitx}
\usepackage{epstopdf}
\usepackage{color}
\usepackage{multirow}
\usepackage{setspace}
\usepackage{overpic}
\usepackage{amssymb}
\usepackage[bookmarksnumbered, pdfstartview=FitH,colorlinks,urlcolor=blue, citecolor=blue,linkcolor=blue] {hyperref}
\usepackage{lineno}
\usepackage{bm}
\usepackage{rotating}
\usepackage{siunitx}
\usepackage{dcolumn}
\usepackage[utf8]{inputenc}
\usepackage{babel}

\let\oldequation\equation
\let\oldendequation\endequation

\renewenvironment{equation}
  {\linenomathNonumbers\oldequation}
  {\oldendequation\endlinenomath}

\begin{document}


\title{\bf \boldmath{
Measurements of absolute branching fractions of $D^{0(+)}\to KKK\pi$ decays}{}
}

\author{
M.~Ablikim$^{1}$\BESIIIorcid{0000-0002-3935-619X},
M.~N.~Achasov$^{4,b}$\BESIIIorcid{0000-0002-9400-8622},
P.~Adlarson$^{77}$\BESIIIorcid{0000-0001-6280-3851},
X.~C.~Ai$^{82}$\BESIIIorcid{0000-0003-3856-2415},
R.~Aliberti$^{36}$\BESIIIorcid{0000-0003-3500-4012},
A.~Amoroso$^{76A,76C}$\BESIIIorcid{0000-0002-3095-8610},
Q.~An$^{73,59,\dagger}$,
Y.~Bai$^{58}$\BESIIIorcid{0000-0001-6593-5665},
O.~Bakina$^{37}$\BESIIIorcid{0009-0005-0719-7461},
Y.~Ban$^{47,g}$\BESIIIorcid{0000-0002-1912-0374}
% ...
\\
\vspace{0.2cm}
(BESIII Collaboration)\\
\vspace{0.2cm} {\it
$^{1}$ Institute of High Energy Physics, Beijing 100049, People's Republic of China\\
% ...
$^{4}$ Budker Institute of Nuclear Physics SB RAS (BINP), Novosibirsk 630090, Russia\\
% ...
$^{36}$ Johannes Gutenberg University of Mainz, Johann-Joachim-Becher-Weg 45, D-55099 Mainz, Germany\\
$^{37}$ Joint Institute for Nuclear Research, 141980 Dubna, Moscow region, Russia\\
% ...
$^{47}$ Peking University, Beijing 100871, People's Republic of China\\
% ...
$^{58}$ Southeast University, Nanjing 211100, People's Republic of China\\
$^{59}$ State Key Laboratory of Particle Detection and Electronics, Beijing 100049, Hefei 230026, People's Republic of China\\
% ...
$^{73}$ University of Science and Technology of China, Hefei 230026, People's Republic of China\\
% ...
$^{76}$ University of Turin and INFN, (A)University of Turin, I-10125, Turin, Italy; (B)University of Eastern Piedmont, I-15121, Alessandria, Italy; (C)INFN, I-10125, Turin, Italy\\
$^{77}$ Uppsala University, Box 516, SE-75120 Uppsala, Sweden\\
% ...
$^{82}$ Zhengzhou University, Zhengzhou 450001, People's Republic of China\\
\vspace{0.2cm}
$^{\dagger}$ Deceased\\
% ...
$^{b}$ Also at the Novosibirsk State University, Novosibirsk, 630090, Russia\\
% ...
$^{g}$ Also at State Key Laboratory of Nuclear Physics and Technology, Peking University, Beijing 100871, People's Republic of China\\
% ...
}
}

\begin{abstract}
Using an $e^+e^-$ sample of $20.3\,\rm fb^{-1}$ collected at the center-of-mass energy $\sqrt{s}=$ 3.773 GeV with the BESIII detector, we report measurements of several four-body hadronic decays of the $D$ mesons.
The absolute branching fractions are determined to be
${\mathcal B}(D^0\to K^0_S K^+K^-\pi^0     )=( 18.4^{+2.6}_{-2.5}\pm 2.4)\times 10^{-5}$,
${\mathcal B}(D^0\to K^0_S K^0_S K^-\pi^+  )=( 12.9^{+1.7}_{-1.6}\pm 2.5)\times 10^{-5}$,
${\mathcal B}(D^0\to K^0_S K^0_S K^+\pi^-)=(5.7^{+1.2}_{-1.1}\pm 1.3)\times 10^{-5}$,
${\mathcal B}(D^0\to K^+K^-K^-\pi^+     )=(17.4^{+1.8}_{-1.7}\pm { 2.2})\times 10^{-5}$, and
${\mathcal B}(D^+\to K^0_S K^+K^-\pi^+)=(13.8^{+2.4}_{-2.2}\pm 2.5)\times 10^{-5}$.
Furthermore, significant $\phi$ signals are found in the decay channels involving $K^+K^-$ pair,
and the corresponding branching fractions are measured as
${\mathcal B}(D^0\to \phi K^0_S\pi^0     )=( 22.7^{+5.4}_{-5.1}\pm 3.7)\times 10^{-5}$,
${\mathcal B}(D^0\to \phi K^-\pi^+     )=(25.2^{+3.5}_{-3.3}\pm 4.6)\times 10^{-5}$,
${\mathcal B}(D^+\to \phi K^0_S\pi^+)=(16.5 ^{+6.0}_{-5.3}\pm 2.6 )\times 10^{-5}$.
The branching fractions of
 $D^0\to K^0_S K^+K^-\pi^0$, 
$D^0\to \phi K^0_S\pi^0$, and
$D^+\to \phi K^0_S \pi^+$ are measured for the first time, 
and those of $D^0\to K^0_S K^0_SK^-\pi^+$, $D^0\to K^0_S K^0_SK^+\pi^-$, $D^0\to K^+K^-K^-\pi^+$, $D^0\to \phi K^-\pi^+$, and $D^+\to K^0_S K^+K^-\pi^+$
are measured with improved precision.
The first uncertainties are statistical and the second are systematic.
\end{abstract}

\newcommand{\BESIIIorcid}[1]{\href{https://orcid.org/#1}{\hspace*{0.1em}\raisebox{-0.45ex}{\includegraphics[width=1em]{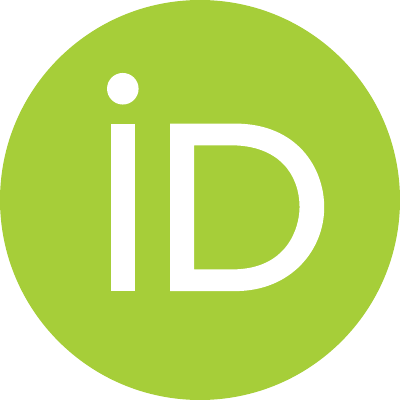}}}}
\maketitle
\section{Introduction}

Hadronic decays of charmed mesons provide a sensitive probe for studying $D^0-\bar D^0$ mixing and $CP$ violation, which are crucial for understanding the imbalance between matter and antimatter in the universe~\cite{matter_antimatter}.

Since the discovery of $D$ mesons in the 1970s, hadronic $D$ decays have been investigated extensively and precisely~\cite{DCS_Panx, DCS_Kaikai}. However, the $D^{0(+)}\to KKK\pi$ decays remain poorly explored. To date, only the FOCUS experiment measured the branching fractions of $D^0\to K^+K^-K^-\pi^+$~\cite{BFKKKpi} and $D^+\to K^0_S K^+K^-\pi^+$~\cite{BFKSKKpi} as well as the combined branching fraction of $D^0\to K^0_S K^0_S K^-\pi^+$ and $D^0\to K^0_S K^0_S K^+\pi^-$~\cite{BFKSKSKpi} relative to $D^0\to K^0_S\pi^+\pi^-$ with large uncertainty. Other multi-body decays, such as 
$D^0\to K^0_S K^+K^-\pi^0$, have not been measured.
Experimentally,  studies of these decays are challenging due to the limited data
samples, high background levels, and low reconstruction efficiencies.

In this paper, we report the measurements of the absolute branching fractions of
$D^0\to K^0_S K^+K^-\pi^0$,
$D^0\to K^0_S K^0_S K^-\pi^+$,
$D^0\to K^0_S K^0_S K^+\pi^-$,
$D^0\to K^+K^-K^-\pi^+$, and
$D^+\to K^0_S K^+K^-\pi^+$
by analyzing $e^+e^-$ collision data corresponding to an integrated luminosity of 20.3~fb$^{-1}$ ~\cite{Dluminosity2024}, collected at the
center-of-mass energy of $\sqrt s=$ 3.773~GeV with the BESIII detector.
In addition, we also measure the branching fractions of the $\phi$-dominated modes
$D^0\to \phi K^0_S\pi^0$, 
$D^0\to \phi K^-\pi^+$, 
and $D^+\to \phi K^0_S\pi^+$.
Throughout this paper, charge conjugate processes are always implied.

\section{BESIII detector and Monte Carlo simulation}

The BESIII detector is a magnetic spectrometer~\cite{BESIII} located at the Beijing Electron
Positron Collider (BEPCII)~\cite{Yu:IPAC2016-TUYA01}. The
cylindrical core of the BESIII detector consists of a helium-based
 multilayer drift chamber (MDC), a plastic scintillator time-of-flight
system (TOF), and a CsI (Tl) electromagnetic calorimeter (EMC),
which are all enclosed in a superconducting solenoidal magnet
providing a 1.0~T magnetic field. 

The solenoid is supported by an octagonal flux-return yoke with resistive plate
counter muon identification modules interleaved with steel.  The
charged-particle momentum resolution at $1~{\rm GeV}/c$ is $0.5\%$, and the specific energy loss $\text{d}E/ \text{d}x$ resolution is $6\%$ for electrons from Bhabha scattering. The EMC measures photon energies with a resolution of $2.5\%$ ($5\%$) at $1$~GeV in the
barrel (end-cap) region. The time resolution in the TOF barrel region is 68~ps,
while that in the end-cap region was 110~ps.  The end-cap TOF system was
upgraded in 2015 using multigap resistive plate chamber technology, providing a
time resolution of 60~ps, which benefits 86\% of the data used in this
analysis~\cite{TOF1,TOF2,TOF3}.

Monte Carlo (MC) simulated data samples, produced with a {\sc geant4}-based~\cite{geant4} software package including the geometric description of the BESIII detector and the
detector response, are used to determine the detection efficiency
and estimate the backgrounds. The simulation includes the beam-energy spread and initial-state radiation in the $e^+e^-$
annihilations modeled with the generator {\sc kkmc}~\cite{kkmc1}.
The inclusive MC samples consist of the production of $D\bar{D}$
pairs, the non-$D\bar{D}$ decays of the $\psi(3770)$, the initial-state radiation
production of the $J/\psi$ and $\psi(3686)$ states, and the
continuum processes.
The known decay modes are modeled with {\sc
evtgen}~\cite{evtgen, Ping:2008zz} using the branching fractions taken from the
Particle Data Group (PDG)~\cite{PDG2022}, and the remaining unknown decays are estimated with {\sc
lundcharm}~\cite{lundcharm,lundcharm2}. The final-state radiation
from charged final-state particles is included in the {\sc
photos} package~\cite{photos}.

The signal MC samples for decays $D^0\to K^0_S K^+K^-\pi^0$,
$D^0\to K^0_S K^0_S K^-\pi^+$,
$D^0\to K^0_S K^0_S K^+\pi^-$,
$D^0\to K^+K^-K^-\pi^+$,
$D^+\to K^0_S K^+K^-\pi^+$,
$D^0\to \phi K^0_S\pi^0$, 
$D^0\to \phi K^-\pi^+$, 
and $D^+\to \phi K^0_S\pi^+$ 
are generated with possible intermediate resonant contributions. 
Each $D^0\to K^0_SK^0_SK\pi$ decay mode is generated with a phase-space~(PHSP) model. For $D\to K^+K^-K\pi$, the signal MC samples are generated as mixtures of $D\to \phi K\pi$, $D\to \phi K^*(892)$, $D\to K^+K^-K^*(892)$, and nonresonant $D\to K^+K^-K\pi$.
For $D\to \phi K\pi$, the MC samples are generated as mixtures of $D\to \phi K\pi$ and $D\to \phi K^*(892)$. 
The mixing ratios for the signal decays $D^0\to K^+K^-K^-\pi^+$ and $D^0\to \phi K^-\pi^+$ are taken from the PDG~\cite{PDG2022}, while those for other decays are determined by studying the invariant mass distributions of the  $K^+K^-$, $K^-\pi^+$, $K^\pm\pi^0$, and $K^0_S\pi^\pm$ in data.

\section{Measurement Method}

The $D^0\bar D^0$ or $D^+D^-$ pairs are produced without any additional hadron in $e^+e^-$ annihilations at $\sqrt s=3.773$~GeV. This process provides a clean environment to measure the branching fractions of hadronic $D$ decays using the double-tag~(DT) method~\cite{tag_double}.
First, single-tag~(ST) candidates are selected by reconstructing a $\bar D^0$ or $D^-$ in the following hadronic final states:
$\bar D^0 \to K^+\pi^-$, $K^+\pi^-\pi^0$, $K^+\pi^-\pi^-\pi^+$, and
$D^- \to K^{+}\pi^{-}\pi^{-}$,
$K^0_{S}\pi^{-}$, $K^{+}\pi^{-}\pi^{-}\pi^{0}$, $K^0_{S}\pi^{-}\pi^{0}$, $K^0_{S}\pi^{+}\pi^{-}\pi^{-}$, $K^{+}K^{-}\pi^{-}$.
Events in which a signal candidate is selected in the presence of an ST $\bar D$ meson
are called DT events.
The branching fraction of the signal decay is determined by
\begin{equation}
\label{eq:br}
{\mathcal B}_{{\rm sig}} = N^{}_{\rm DT}/(N^{\rm tot}_{\rm ST}\cdot\epsilon_{{\rm sig}}),
\end{equation}
where $N^{\rm tot}_{\rm ST}=\sum_i N_{{\rm ST}}^i$ and $N^{}_{\rm DT}$ are the total ST yield and the signal yield in the data, respectively, and $N_{{\rm ST}}^i$ is the ST yield of the tag mode $i$.
The efficiency $\epsilon_{{\rm sig}}$ of detecting the signal $D$ decay, averaged over all tag modes $i$, is given by
\begin{equation}
\label{eq:eff}
\epsilon_{{\rm sig}} = \sum_i (N^i_{{\rm ST}}\cdot\epsilon^i_{{\rm DT}}/\epsilon^i_{{\rm ST}})/N^{\rm tot}_{\rm ST},
\end{equation}
where $\epsilon^i_{{\rm ST}}$ and $\epsilon^i_{{\rm DT}}$ are the efficiencies of detecting ST and DT candidates in the tag mode $i$, respectively.

The measured branching fractions of neutral $D$ decays need to be corrected for quantum correlation (QC) effects which are present in the data but are not implemented in the MC simulation. Due to limited signal yields in this analysis, the QC effect for $D^0$ decays is considered as a source of systematic uncertainty~\cite{strongphase1,strongphase2}. 

\section{Event selection}
\label{sec:selection}
The selection criteria for $K^\pm$, $\pi^\pm$, $K^0_S$, and $\pi^0$ are adopted from Ref.~\cite{TaoLY}.
All charged tracks, except those originating from $K^0_{S}$ decays, are required to have a polar angle $\theta$ with respect to the symmetry axis of the MDC within the MDC acceptance, $|\rm{\cos\theta}|<0.93$,  and a distance of closest approach to the interaction point~(IP) within 10~cm along the beam direction
and within 1~cm in the plane perpendicular to the beam direction. Particle identification~(PID) for charged pions and kaons is performed by
using TOF information and the $\mathrm d E/\mathrm d x$ measured by the MDC to calculate
confidence levels for the pion and kaon hypotheses ($CL_{\pi}$ and $CL_{K}$). Kaon and pion candidates are identified
via $CL_{K}>CL_{\pi}$ and $CL_{\pi}>CL_{K}$, respectively.

The $K^0_S$ candidates are formed with two oppositely charged tracks, which are assigned the pion hypothesis without applying PID criteria.
Due to the long lifetime of the $K^0_S$  meson,
the distance of closest approach to the IP in the beam direction must be less than 20~cm, and there is no requirement for the distance of closest approach in the plane transverse to the beam direction. The two charged tracks are constrained to originate from a common vertex with a flight distance from the IP at least twice the vertex resolution. To improve the mass resolution, the primary vertex fit and secondary vertex fit must satisfy $\chi^2<100$. The invariant mass of $\pi^+\pi^-$ is required to be within $(0.487,0.511)~{\rm GeV}/c^2$. 

The $\pi^0$ candidates are formed with photon pairs. The photon candidates are identified using isolated showers in the EMC.
The deposited energy of each shower must be more than 25~MeV in the barrel region ($|\cos \theta|< 0.80$) and more than 50~MeV in the end-cap region ($0.86 <|\cos \theta|< 0.92$).  
To exclude showers that originate from charged tracks,
the angle subtended by the EMC shower and the position of the closest charged track at the EMC
must be greater than 10 degrees as measured from the IP. 
To suppress electronic noise and showers unrelated to the event, the difference between the EMC time and the event start time is required to be within 
[0, 700]\,ns.
Photon pairs with an invariant mass in the range of $(0.115,\,0.150)$\,GeV$/c^{2}$ are taken as $\pi^0$ candidates. To improve the mass resolution, a kinematic fit constraining the invariant mass of $\gamma\gamma$ to the nominal $\pi^0$ mass~\cite{PDG2022} is applied, and the resulting $\chi^2$ of the fit is required to be less than 50.

In the selection of $\bar D^0\to K^+\pi^-$ tag candidates, backgrounds from cosmic rays and Bhabha events are rejected with the same requirements as in Ref.~\cite{deltakpi}.
The TOF time difference between the two charged tracks is required to be less than 5 ns. The track pair must not be consistent with being a muon or electron-positron pair. Additionally, there must be at least one EMC shower with a deposited energy of more than 50~MeV, or at least one additional charged track detected in the MDC.

\section{Yields of ST $\bar D$ mesons}
\label{sec:singletag}

The tagged $\bar D$ mesons are identified using two variables, namely the energy difference
\begin{equation}
\Delta E_{\rm tag} \equiv E_{\rm tag} - E_{\rm beam},
\label{eq:deltaE}
\end{equation}
and the beam-constrained mass
\begin{equation}
M_{\rm BC}^{\rm tag} \equiv \sqrt{E^{2}_{\rm beam}/c^{4}-|\vec{p}_{\rm tag}|^{2}/c^{2}}.
\label{eq:mBC}
\end{equation}
Here, $E_{\rm beam}$ is the beam energy, and
$\vec{p}_{\rm tag}$ and $E_{\rm tag}$ are the momentum and the energy of
the $\bar D$ candidate in the rest frame of the $e^+e^-$ system, respectively.
For each tag mode, if there are multiple candidates in an event,
only the one with the minimum $|\Delta E_{\rm tag}|$ is kept.
The tagged $\bar D$ candidates must fulfill mode-dependent
$\Delta E_{\rm tag}$ requirements due to different resolutions, as listed in the second columns of Tables~\ref{tab:D0tagDE} and~\ref{tab:DptagDE}.

\begin{figure*}[htp]
  \centering
  \setlength{\abovecaptionskip}{1pt}
  \setlength{\belowcaptionskip}{1pt}
\includegraphics[width=0.75\linewidth]{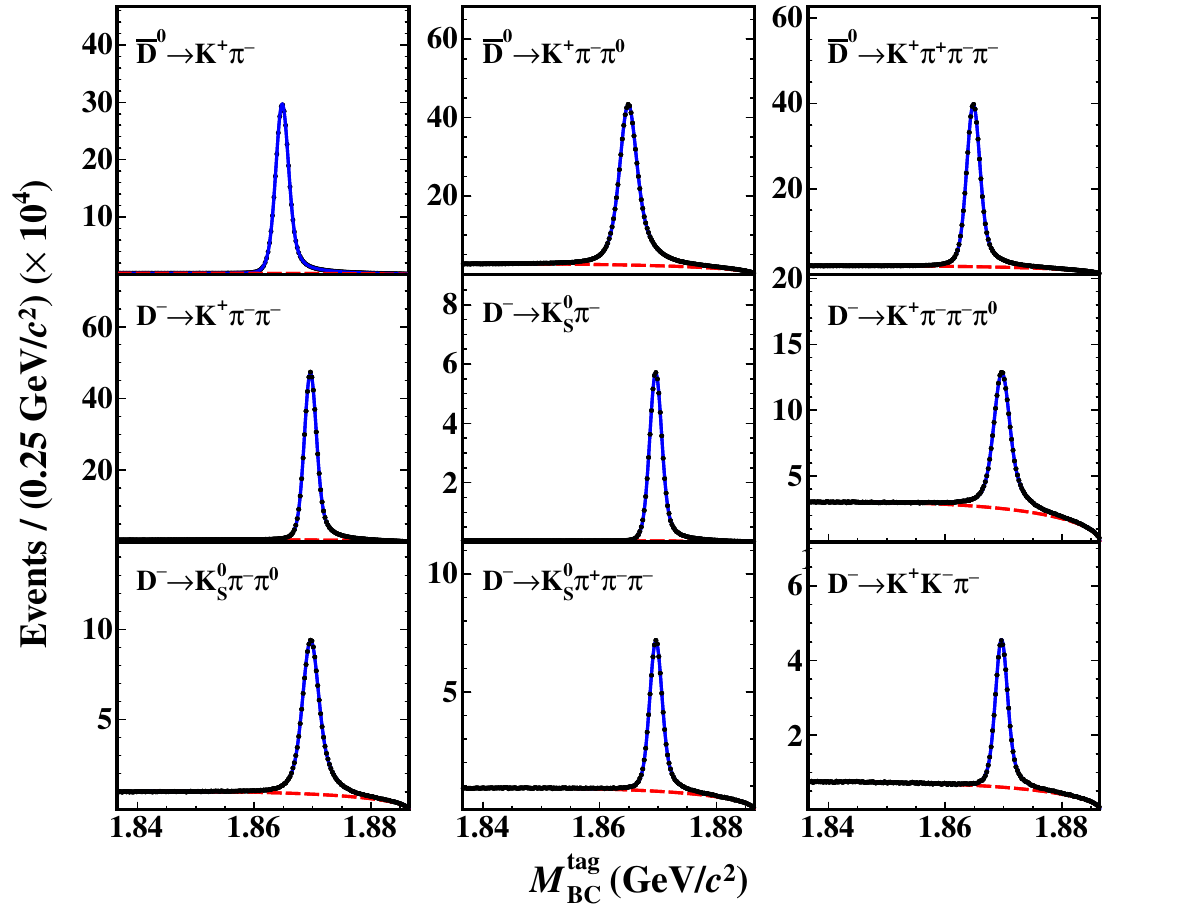}
  \caption{\small
Fits of the ST $\bar D^0$ and $D^{-}$ candidates to the $M_{\rm BC}^{\rm tag}$ distributions.
The dots with error bars show the data,
the blue solid curves are the fit results and the red dotted lines are
the fitted background shapes.}
\label{fig:datafit_MassBC}
\end{figure*}

To determine the ST yields of  $\bar D$ mesons for individual tag modes, unbinned maximum-likelihood fits are performed on the $M_{\rm BC}^{\rm tag}$
distributions of the ST candidates, following the procedure described in Ref.~\cite{TaoLY}.
In the fits, the $\bar D$ signal is modeled by an MC-simulated shape convolved with
a Gaussian function with free parameters accounting for the resolution difference between data and MC simulation.
The resulting fits on the $M_{\rm BC}^{\rm tag}$
distributions for different tag modes are shown in
Fig.~\ref{fig:datafit_MassBC}. The extracted ST  $\bar D^0$ and $D^-$ yields in data and the corresponding ST efficiencies are summarized in the third and fourth columns of Tables~\ref{tab:D0tagDE} and~\ref{tab:DptagDE}. 
The total yields of the ST $\bar D^0$  and $D^-$ mesons in data are $(1688.7\pm0.5)\times 10^4$ and
$(1096.0\pm0.4)\times 10^4$, respectively, where the uncertainties are statistical only.

\begin{table*}[htbp]
\centering

\caption{The $\Delta E_{\rm tag}$ requirements, ST yields~($N_{\rm ST}$), ST efficiencies~($\epsilon_{\rm ST}$) and DT efficiencies~($\epsilon_{\rm DT}$) for $D^0$ decays. The number 1-6 represents the signal decays: $^1D^0\to K^0_S K^+K^-\pi^0$,
$^2D^0\to \phi K^0_S\pi^0$,
$^3D^0\to K^0_S K^0_S K^-\pi^+$,
$^4D^0\to K^0_S K^0_S K^+\pi^-$,
$^5D^0\to K^+K^-K^-\pi^+$ and
$^6D^0\to \phi K^-\pi^+$. The uncertainties are statistical only.}

\label{tab:D0tagDE}
\resizebox{1.0\textwidth}{!}{
\begin{tabular}{lcrccccccc}
\hline \hline \multirow{2}{*}{Tag mode} & $\Delta E_{\rm tag}$ & $N_{\text{ST }}$  & $\epsilon_{\rm ST}$&$\epsilon_{\rm DT}^{1}$& $\epsilon_{\rm DT}^{2}$& $\epsilon_{\rm DT}^{3}$& $\epsilon_{\rm DT}^{4}$& $\epsilon_{\rm DT}^{5}$& $\epsilon_{\rm DT}^{6}$ \\

&$(\mathrm{MeV})$&($\times10^4$)&$(\%)$&$(\%)$&$(\%)$&$(\%)$&$(\%)$&$(\%)$&$(\%)$\\\hline
$\bar{D}^0 \rightarrow K^{+} \pi^{-}$& $(-27,27)$  & $382.1 \pm 0.2$ & $66.67\pm 0.01$ &$2.84\pm0.02$& $2.11\pm0.02$& $4.56\pm0.05$& $4.74\pm0.05$& $3.29\pm0.02$& $3.16\pm0.03$\\
$\bar{D}^0 \rightarrow K^{+} \pi^{-} \pi^0$& $(-62,49)$  & $792.6 \pm 0.3$ & $37.92\pm 0.01$&$1.42\pm0.02$& $1.06\pm0.02$& $2.46\pm0.03$& $2.34\pm0.03$& $1.97\pm0.02$& $1.90\pm0.02$ \\
$\bar{D}^0 \rightarrow K^{+} \pi^{-} \pi^{-} \pi^{+}$ &$(-26,24)$ & $514.0 \pm 0.2$ & $42.20\pm 0.01 $&$1.50\pm0.02$& $1.15\pm0.02$& $2.43\pm0.03$& $2.33\pm0.03$& $1.88\pm0.02$& $1.81\pm0.02$ \\

\hline \hline
\end{tabular}
}

\end{table*}

\begin{table*}[htbp]
\centering

\caption{The $\Delta E_{\rm tag}$ requirements, ST yields~($N_{\rm ST}$), ST efficiencies~($\epsilon_{\rm ST}$) and DT efficiencies~($\epsilon_{\rm DT}$) for $D^+$ decays.  The number 7-8 presents the signal decays: $^7D^+\to K^0_S K^+K^-\pi^+$
and
$^8D^+\to \phi K^0_S\pi^+$. The uncertainties are only statistical.}

\label{tab:DptagDE}
\begin{tabular}{lcrccc}
\hline \hline \multirow{1}{*}{Tag mode} &  $\Delta E_{\rm tag}~(\mathrm{MeV})$ & $N_{\text{ST }}~(\times 10^4)$  & $\epsilon_{\rm ST}~(\%)$ & $\epsilon_{\rm DT}^{7}~(\%)$ & $\epsilon_{\rm DT}^{8}~(\%)$\\\hline

$D^{-} \rightarrow K^{+} \pi^{-} \pi^{-}$ &$(-25,24)$ & $567.4 \pm 0.2$  & $52.40 \pm 0.01$ & $2.78\pm0.02$& $1.86\pm0.02$\\
$D^{-} \rightarrow K_S^0 \pi^{-}$&$(-25,26)$ & $66.7 \pm 0.1$  & $52.60 \pm 0.01$& $2.75\pm0.02$& $1.81\pm0.02$ \\
$D^{-} \rightarrow K^{+} \pi^{-} \pi^{-} \pi^0$&$(-57,46)$  & $181.0 \pm 0.2 $ & $25.62 \pm 0.01$ & $1.17\pm0.01$& $0.75\pm0.01$\\
$D^{-} \rightarrow K_S^0 \pi^{-} \pi^0$ &$(-62,49)$ & $150.7 \pm 0.1$ & $27.74 \pm 0.01$ & $1.39\pm0.02$& $0.95\pm0.02$\\
$D^{-} \rightarrow K_S^0 \pi^{-} \pi^{-} \pi^{+}$&$(-28,27)$  & $81.0 \pm 0.1$ & $30.25\pm 0.01$& $1.30\pm0.01$& $0.86\pm0.01$ \\
$D^{-} \rightarrow K^{+} K^{-} \pi^{-}$& $(-24,23)$  & $49.2 \pm 0.1$ & $41.94\pm 0.01$& $2.17\pm0.02$& $1.42\pm0.02$ \\
\hline
\hline
\end{tabular}
\end{table*}

\section{Yields of DT events}
\label{sec:DTyield}

Candidates for the signal decays $D^0\to K^0_S K^+K^-\pi^0$,
$D^0\to K^0_S K^0_S K^-\pi^+$,
$D^0\to K^0_S K^0_S K^+\pi^-$,
$D^0\to K^+K^-K^-\pi^+$, and
$D^+\to K^0_S K^+K^-\pi^+$ are reconstructed from the remaining charged tracks and neutral showers that are not used in the reconstruction of the ST $\bar D$ candidates.

The signal $D$ mesons are identified using the energy difference $\Delta E_{\rm sig}$
and the beam-constrained mass $M_{\rm BC}^{\rm sig}$, which are calculated with the ``sig'' analogues of the ``tag'' in Eqs.~(\ref{eq:deltaE}) and (\ref{eq:mBC}), respectively.
For each signal mode, if there are multiple candidates in an event, only the one with the minimum $|\Delta E_{\rm sig}|$ is kept.
The signal decays are required to satisfy the mode-dependent $\Delta E_{\rm sig}$ requirements,
as shown in the second column of Table~\ref{tab:DT}.

\begin{figure*}[htbp]
  \centering
    \setlength{\abovecaptionskip}{1pt}
  \setlength{\belowcaptionskip}{1pt}
\includegraphics[width=0.75\linewidth]{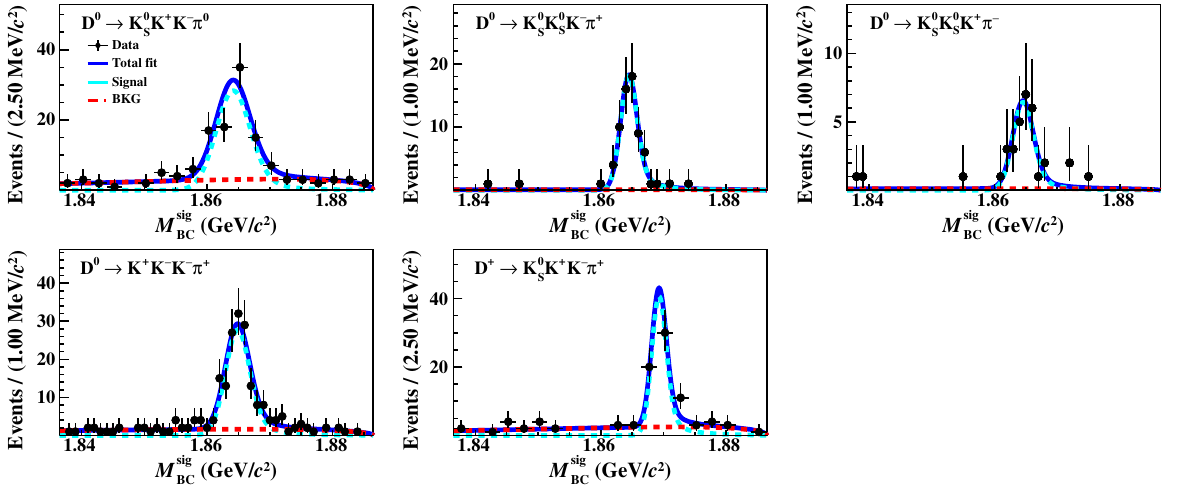}
  \caption{\small
The fits of the DT candidate events to the $M^{\rm sig}_{\rm BC}$ distributions for $D^0\to K^0_S K^+K^-\pi^0$,
$D^0\to K^0_S K^0_S K^-\pi^+$,
$D^0\to K^0_S K^0_S K^+\pi^-$,
$D^0\to K^+K^-K^-\pi^+$, and
$D^+\to K^0_S K^+K^-\pi^+$. The dots with error bars are data, the blue solid curves are the total fit shapes, the cyan dashed curves show fits to the signal shapes
and the red dashed curves fits to the background shapes.
}
\label{fig:2Dfit1}
\end{figure*}

The yield of each signal decay ($N^{\rm fit}_{\rm DT}$) is obtained from an unbinned maximum-likelihood
fit to the $M_{\rm BC}^{\rm sig}$ distribution of the selected DT candidates. In the fit, the signal shape  is modeled by an MC-simulated shape convolved with
a Gaussian function, while the combinatorial background is described by an ARGUS function~\cite{ARGUS}.
Here, the parameters of the convolved Gaussian resolution function and the ARGUS function are free.
The signal decays involving a $K^0_S$ in the final state suffer from non-resonant $\pi^+\pi^-$ background surviving the $K^0_S$ selection criteria. Such background events may form peaking backgrounds around the nominal $D^0$ or $D^{+}$ mass in the $M_{\rm BC}^{\rm sig}$ distributions. This background contribution is estimated with events in the $K^0_S$ sideband region of
$0.020<|M_{\pi^+\pi^-}-0.498|<0.044~{\rm GeV}/c^2$.
No significant peaking background is observed in any signal decay, 
and it is neglected in the nominal measurements, but treated as a source of systematic uncertainty.
The resulting fits for the five decays 
$D^0\to K^0_S K^+K^-\pi^0$,
$D^0\to K^0_S K^0_S K^-\pi^+$,
$D^0\to K^0_S K^0_S K^+\pi^-$,
$D^0\to K^+K^-K^-\pi^+$, and
$D^+\to K^0_S K^+K^-\pi^+$ are shown in Fig.~\ref{fig:2Dfit1}.
For each signal decay, the statistical significance is greater than $5\sigma$,
which is calculated as $\sqrt{2\times( {\rm ln}\mathcal{L}_{\rm sig}-{\rm ln}\mathcal{L}_{0})}$,
where ${\rm ln}\mathcal{L}_{\rm sig}$ 
and ${\rm ln}\mathcal{L}_{0}$ are the natural logarithm of likelihoods 
with and without the signal component in the fit~\cite{significance_origin}.
The obtained fitted signal yields of $D\to KKK\pi$ decays and their statistical significance are listed in the third and fourth columns of Table~\ref{tab:DT}.

\begin{figure*}[htbp]
  \centering
    \setlength{\abovecaptionskip}{1pt}
  \setlength{\belowcaptionskip}{1pt}
\includegraphics[width=0.75\linewidth]{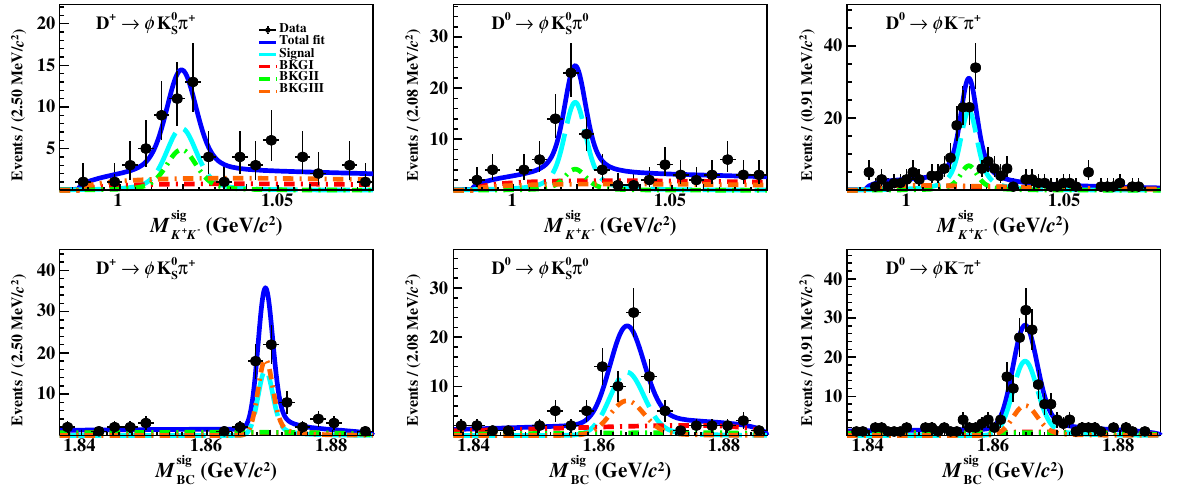}
  \caption{\small
Fits to the $M_{\rm BC}^{\rm sig}$ versus $M^{\rm sig}_{ K^+K^-}$ distributions of DT events for
$D^0\to \phi K^0_S \pi^0$, $D^0\to \phi K^-\pi^+$, and $D^+\to \phi K^0_S \pi^+$.
The dots with error bars are data, cyan dashed curves are fit signal, red dashed curves are fit BKGI, green dashed curves are fit BKGII and the orange dashed curves are fit BKGIII.
}
\label{fig:2Dfit2}
\end{figure*}

We further examine the $M_{K^+K^-}$ distributions of the accepted DT candidates
for $D^0\to K^0_S K^+K^-\pi^0$, $D^0\to K^+K^-K^-\pi^+$, and $D^+\to K^0_S K^+K^-\pi^+$,
and significant $\phi$ signals are found.
To determine the signal yield of each $D\to \phi K\pi$ decay, we perform a two-dimensional unbinned maximum-likelihood fit to
the $M_{K^+K^-}$ versus $M_{\rm BC}^{\rm sig}$ distribution.
In the fit, the $\phi$ signal is described by a Breit-Wigner~(BW) function convolved with a Gaussian 
 function, while
the combinatorial $K^+K^-$ background is described by a reversed ARGUS function, i.e. an ARGUS function where the independent variable $m$ is replaced by $2m_0-m$.
The mass and width of $\phi$ are fixed to their PDG values and the other parameters are allowed to float. There are four components in the two-dimensional fit. The first is the signal part~($S_{M_{\rm BC}^{\rm sig}}\cdot S_{M_{K^+K^-}}$), the second is BKGI~($S_{M_{\rm BC}^{\rm sig}}\cdot B_{M_{K^+K^-}}$), the third is BKGII~($B_{M_{\rm BC}^{\rm sig}}\cdot S_{M_{K^+K^-}}$), the last is BKGIII~($B_{M_{\rm BC}^{\rm sig}}\cdot B_{M_{K^+K^-}}$), where $S_{M_{\rm BC}^{\rm sig}}$ is the ${\rm MC~shape(\it M_{\rm BC}^{\rm sig})}\otimes {\rm Gaussian}$ ($\otimes$ represents convolution), $S_{M_{K^+K^-}}$ is ${\rm BW\otimes Gaussian}$, $B_{M_{\rm BC}^{\rm sig}}$ is an ARGUS function and $B_{M_{K^+K^-}}$ is a reversed ARGUS function.
The obtained fit results for these three decays are shown in Fig.~\ref{fig:2Dfit2}.
The same method as above is also used to evaluate the statistical significance of each $D\to \phi K\pi$ decay.
The obtained fitted signal yields of $D\to \phi K\pi$ decays and their statistical significance are also listed in the third and fourth columns of Table~\ref{tab:DT}.

\begin{figure*}[htbp]
  \centering
    \setlength{\abovecaptionskip}{1pt}
  \setlength{\belowcaptionskip}{1pt}
\includegraphics[width=1.0\linewidth]{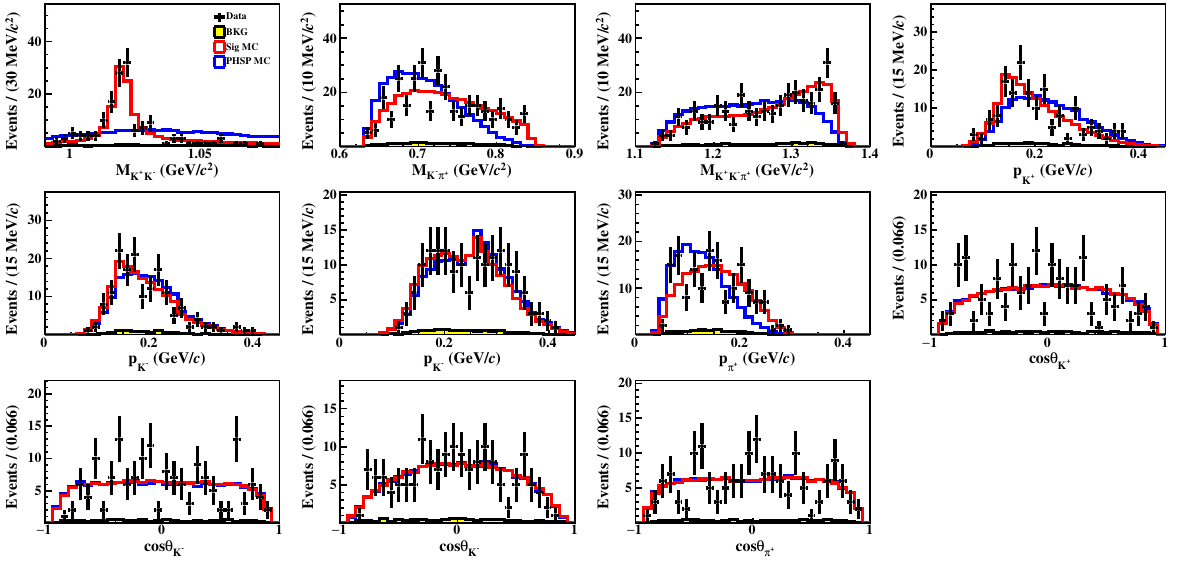}

  \caption{
  Distributions of the momenta and cosines of polar angles of the daughter particles as well as the invariant masses of two-body and three-body combinations of the $D^+\to K^+K^-K^-\pi^+$ candidate events, comparing
  data (dots with error bars) and the signal MC samples (red and blue histograms). The red histogram shows the mixed signal MC sample and the blue one the PHSP MC. The MC-simulated backgrounds~(BKG) from the inclusive MC sample are shown as yellow filled histograms.}
\label{fig:compare}

\end{figure*}

\begin{table*}[htbp]
\centering
\caption{ \small
Requirements on $\Delta E_{\rm sig}$, fitted signal yields in data ($N_{\rm DT}$), signal efficiencies which
include the branching fractions of the sub-decays $K^0_S\to \pi^+\pi^-$, $\pi^0\to \gamma\gamma$ or $\phi$ ($\epsilon_{\rm sig}$),
statistical significance~($\mathcal S$),
and the obtained branching fractions~(${\mathcal B}_{\rm sig}$) for the different signal decays.
The first and second uncertainties for ${\mathcal B}_{\rm sig}$ are statistical and systematic, respectively,
while the uncertainties for $N_{\rm DT}$ and $\epsilon_{\rm sig}$ are statistical only.
For comparison, the relevant PDG branching fractions~($\mathcal{B}_{\rm PDG}$) are listed in the last column.}
\label{tab:DT}
\begin{tabular}{l c c c c r l}
  \hline\hline

  {Signal mode}
  &{ $\Delta E_{ \rm sig}$~(MeV)}
  &{$N_{\rm DT}$}
  &{ $\mathcal S~(\sigma)$}
  &$\epsilon_{\rm sig}$~(MeV)
  &  \multicolumn{1}{c}{${\mathcal B}_{\rm sig}$}
  & \multicolumn{1}{c}{${\mathcal B}_{\rm PDG}$}\\\hline

$D^0\to K^0_S K^+K^-\pi^0$ &$(-27,23)$&81.0$^{+11.3}_{-10.7}$&$>10$ &2.60$\pm0.02$&$18.4^{+2.6}_{-2.5}\pm 2.4$&\multicolumn{1}{c}{...}\\

$D^0\to \phi K^0_S \pi^0$ &$(-27,23)$ &36.8$^{+8.7}_{-8.1}$&6.4&0.96$\pm0.01$&$22.7^{+5.4}_{-5.1}\pm 3.7$&\multicolumn{1}{c}{...}\\ \hline

$D^0\to K^0_S K^0_S K^-\pi^+$ &$(-20,19)$          & 66.2$^{+8.6}_{-7.9}$&$>10$ &3.04$\pm0.04$& $12.9^{+1.7}_{-1.6}\pm 2.5$&\multirow{2}{*}{59.0$\pm13.0$}\\ \cline{1-6}

$D^0\to K^0_S K^0_S K^+\pi^-$&$(-20,19)$ &28.7$^{+5.9}_{-5.3}$&8.7&2.96$\pm0.04$&$5.7^{+1.2}_{-1.1}\pm 1.3$&\\  \hline

$D^0\to K^+K^-K^-\pi^+$    &$(-21,19)$&144.3$^{+14.2}_{-13.5}$&$>10$&4.91$\pm0.04$&$17.4^{+1.8}_{-1.7}\pm { 2.2}$&$22.5\pm3.2$\\

$D^0\to \phi K^-\pi^+$  &$(-21,19)$&98.8$^{+13.5}_{-12.7}$&$>10$&2.32$\pm0.01$&$25.2^{+3.5}_{-3.3}\pm 4.6$&\multicolumn{1}{c}{$22.0\pm4.3$}\\  \hline

$D^+\to K^0_S K^+K^-\pi^+$&$(-21,20)$&52.9$^{+9.0}_{-8.2}$&$9.9$&3.50$\pm0.02$&$13.8^{+2.4}_{-2.2}\pm 2.5$&$24.0\pm 5.0$\\

$D^+\to \phi K^0_S\pi^+$&$(-21,20)$&20.6$^{+7.5}_{-6.6}$&5.0&1.14$\pm0.01$ &$16.5 ^{+6.0}_{-5.3}\pm 2.6$&\multicolumn{1}{c}{...}\\
  \hline\hline
\end{tabular}

\end{table*}

The DT efficiencies are determined from MC simulation,  which are summarized in  Tables~\ref{tab:D0tagDE} and~\ref{tab:DptagDE}, respectively, and the resulting  signal efficiencies averaged over tag modes are summarized in the fifth column of Table~\ref{tab:DT}.
As an example, Fig.~\ref{fig:compare} shows
the distributions of the momenta and cosines of the polar angles of the daughter particles as well as the
invariant masses of two-body or three-body particle combinations of
the candidates for $D^0\to K^+K^-K^-\pi^+$ between data and MC simulation. The comparisons for other $D\to KKK\pi$ decays can be found in the Appendix.

The obtained branching fractions of each signal decay 
are summarized in the sixth column of Table~\ref{tab:DT}.

\section{Systematic uncertainties}
\label{sec:sys}

The systematic uncertainties are estimated relative to the measured branching fractions and discussed below.

The systematic uncertainties in the total yields of ST $\bar D$ mesons, which are mainly due to the fits to the $M_{\rm BC}$ distributions of the ST $\bar D$ candidates, were previously estimated to be
0.3\% for ST $\bar D^0$ and $D^-$~\cite{TaoLY}.

The tracking and PID efficiencies of charged particles are studied with control samples from the DT candidates of
$D^0 \to K^-\pi^+$, $K^-\pi^+\pi^0$,
$K^-\pi^+\pi^+\pi^-$ versus $\bar D^0 \to K^+\pi^-$, $K^+\pi^-\pi^0$,
$K^+\pi^-\pi^-\pi^+$, as well as $D^+ \to K^- \pi^+\pi^+$ versus $D^-\to
K^+\pi^-\pi^-$, which will miss a $\pi^\pm$ or $K^\pm$.  The systematic uncertainties of tracking and PID are assigned as  0.5\% per $K^\pm$ or $\pi^\pm$.

The systematic uncertainty in $K_{S}^{0}$ reconstruction
is estimated using the control samples of
$J/\psi\to K^{*}(892)^{\mp}K^{\pm}$ and $J/\psi\to \phi K_S^{0}K^{\pm}\pi^{\mp}$~\cite{Ksuncer}
and found to be 1.5\% per $K^0_S$.
The systematic uncertainty of $\pi^0$ reconstruction is assigned as 3.0\% per $\pi^0$ with the control sample of $D^0\to K^-\pi^+\pi^0$ versus 
$\bar D^0\to K^+\pi^-$ and $\bar D^0\to K^+\pi^-\pi^-\pi^+$~\cite{TaoLY}.

The uncertainties of the quoted branching fractions of $\pi^0\to \gamma\gamma$, $K^0_S\to \pi^+\pi^-$, and $\phi\to K^+K^-$ are 0.03\%, 0.07\%, and 1.0\%, respectively~\cite{PDG2022}.

The systematic uncertainties due to $\Delta E_{\rm sig}$ requirements are studied using the DT events $D^0\to K_S^0 K^-\pi^+\pi^0$, $D^0\to K_S^0K_S^0\pi^+\pi^-$, $D^0\to K^+K^-\pi^+\pi^-$, $D^+\to K_S^0 K^+\pi^+\pi^-$, and $D^+\to K_S^0K_S^0\pi^+\pi^0$  versus the same $\bar D$ tags in our nominal analysis. The efficiency differences between data and MC simulation, (0.8-3.5)\%, are assigned as the systematic uncertainties for various signal decays.

The systematic uncertainty in the fit to the $M_{\rm BC}^{\rm sig}$ distribution is evaluated by changing the signal and background parameterizations.
Specifically, the signal shape is replaced with a version that excludes Gaussian smearing and the endpoint of the ARGUS function is varied by~($\pm0.2$\,MeV/$c^2$).
To estimate the uncertainty due to the $M_{ K^+K^-}$ fit,
the alternative signal shape is obtained by replacing the original shape with the MC shape and
the uncertainty of the background shape is obtained by varying the endpoint of the inverse-ARGUS function~($\pm0.2$\,MeV/$c^2$).
The relative changes of the signal shape and background shape are assigned as individual systematic uncertainties.
Adding these two effects in quadrature results in systematic uncertainties of (1.4-10.0)\%.

\begin{table*}[htbp]
\centering
\caption{
Relative systematic uncertainties (\%) in the measurements of the
branching fractions of
$^1D^0\to K^0_S K^+K^-\pi^0$,
$^2D^0\to \phi K^0_S\pi^0$,
$^3D^0\to K^0_S K^0_S K^-\pi^+$,
$^4D^0\to K^0_S K^0_S K^+\pi^-$,
$^5D^0\to K^+K^-K^-\pi^+$,
$^6D^0\to \phi K^-\pi^+$,
$^7D^+\to K^0_S K^+K^-\pi^+$, 
and
$^8D^+\to \phi K^0_S\pi^+$.
}

\label{tab:relsysuncertainties}
\centering
\begin{tabular}{|c|S[table-format=2.1]|S[table-format=2.1]|S[table-format=2.1]|S[table-format=2.1]|S[table-format=2.1]|S[table-format=2.1]|S[table-format=2.1]|S[table-format=2.1]|}
\hline\hline
\multicolumn{1}{|c|}{\multirow{2}{*}{Source}}& \multicolumn{6}{c|}{$D^0$}& \multicolumn{2}{c|}{$D^+$} \\ \cline{2-9}
&1&2&3&4&5&6&7&8 \\ \hline
$N^{\rm tot}_{\rm ST}$ & 0.3 &0.3 &0.3 &0.3&0.3 &0.3 &0.3 &0.3 \\
$\pi^\pm$/$K^\pm$ tracking & { 2.0}&2.0 &{ 3.0} &{ 3.0} &2.0 &2.0  &{ 2.5} & 2.5 \\
$\pi^\pm$/$K^\pm$ PID & { 1.0}&1.0 &{ 1.0} &{ 1.0} &2.0 &2.0  &{ 1.5} & 1.5 \\
$K_S^0$ reconstruction & 1.5 &1.5 &3.0 &3.0 & {--}  & {--} &1.5 &1.5 \\
$\pi^0$ reconstruction & 3.0 &3.0 &{--} &{--} &{--} &{--} &{--} &{--}\\
Quoted $\mathcal B$ & 0.1& 1.0&0.2&0.2&{--}&1.0 &0.1&1.0 \\
$\Delta E_{\rm sig}$ cut & 0.8 &0.8 & 0.6 & 0.6 & 3.5 &3.5 & 3.1&3.1 \\
DT fit   &1.5 &3.4 &1.9 &1.4 &4.7& 8.4 &8.9&10.0 \\
$K_S^0$ sideband &2.1 &2.1 &1.5 &0.0 &{--} &{--}  &1.5 &1.5   \\
MC statistics & 1.1 &1.1 &1.0 &1.0 &1.0 &1.0  &1.1 &1.1 \\
MC generator & 9.5 &13.0 &{16.8} &{19.3} &8.1 &13.6  &14.3 & 10.9 \\
QC effect &7.1 &7.1 &7.1  &7.1 &7.1 &7.1 & {--} & {--} \\
\hline
Total  &{ 12.9} &{16.0} &{ 19.0} &{ 21.1} &{ 12.6} & {18.1}&{ 17.6}& {15.7} \\
\hline\hline
\end{tabular}

\end{table*}

The systematic uncertainties due to the $K^0_S$ sideband are examined with and without taking into account its contribution.

The uncertainties due to limited MC statistics are assigned by
$\frac{1}{\sqrt{N}}\sqrt{\frac{1-\varepsilon}{\varepsilon}}$, 
where $\epsilon$ is the detection efficiency and $N$ is the total number of produced signal MC events. For various signal decays, (1.0-1.1)\%, are taken as the systematic uncertainties.

The imperfect simulations of the momentum and $\cos \theta$ distributions of charged particles 
are considered as a source of systematic uncertainty named MC generator. This systematic uncertainty is evaluated by varying the input branching fractions of the known decays within their $\pm 1 \sigma$ ranges and assigning one quarter of this variation for the unknown decays.
The resulting changes in signal efficiencies, (8.1-19.3)\%, are taken as the systematic uncertainties.

The QC effect is corrected by a factor $f_{\rm QC}=\frac{1}{1-C_f(2f_{CP+}-1)}$ as described in Ref.~\cite{QCuncer02}.
For all $D^0$ signal decays, a conservative uncertainty, 7.1\%, due to QC effects is assigned as the systematic uncertainty~\cite{strongphase1,strongphase2}.

Table~\ref{tab:relsysuncertainties} summarizes the systematic uncertainties in the branching fraction measurements. For each signal decay, the total systematic uncertainty is obtained by adding all individual contributions in quadrature. The obtained total systematic uncertainties range between~(12.2-21.3)\% for different signal decays.

\section{Summary}

By analyzing $20.3\,\rm fb^{-1}$ of $e^+e^-$ collision data  taken at $\sqrt{s}=$ 3.773 GeV with the BESIII detector,
we determine the branching fractions of the hadronic decays
$D^0\to K^0_S K^+K^-\pi^0$,
$D^0\to K^0_S K^0_S K^-\pi^+$,
$D^0\to K^0_S K^0_S K^+\pi^-$,
$D^0\to K^+K^-K^-\pi^+$, and
$D^+\to K^0_S K^+K^-\pi^+$.
The decays $D^0\to \phi K^0_S\pi^0$,
$D^0\to \phi K^-\pi^+$, and $D^+\to \phi K^0_S\pi^+$ are
found to be the main sub-processes in
$D^0\to K^0_S K^+K^-\pi^0$, $D^0\to K^+K^-K^-\pi^+$, and $D^+\to K^0_S K^+K^-\pi^+$,
respectively; and their branching fractions have been determined.
All the obtained results are shown in Table~\ref{tab:DT}.

The branching fractions of $D^0\to K^0_S K^+K^-\pi^0$, 
$D^0\to \phi K^0_S\pi^0$, and $D^+\to \phi K^0_S \pi^+$ are measured for the first time; 
and those of $D^0\to K^0_S K^0_SK^-\pi^+$, $D^0\to K^0_S K^0_SK^+\pi^-$, $D^0\to K^+K^-K^-\pi^+$, $D^0\to \phi K^-\pi^+$, and $D^+\to K^0_S K^+K^-\pi^+$
are measured with improved precision.
Our branching fractions of $D^0\to K^+K^-K^-\pi^+$ and $D^+\to K^0_S K^+K^-\pi^+$
are consistent with the respective PDG values~\cite{PDG2022} within $1.2\sigma$ and 1.7$\sigma$, respectively.
However, the summed branching fraction of $D^0\to K^0_S K^0_S K^-\pi^+$ and $D^0\to K^0_S K^0_S K^+\pi^-$ presented in this work deviates
from the combined branching fraction of $D^0\to K^0_S K^0_S K^\pm\pi^\mp$ based on the FOCUS measurement
by $3.0\sigma$.

\section{Acknowledgement}

The BESIII Collaboration thanks the staff of BEPCII (https://cstr.cn/31109.02.BEPC) and the IHEP computing center for their strong support. This work is supported in part by National Key R\&D Program of China under Contracts Nos. 2023YFA1606000, 2023YFA1606704; National Natural Science Foundation of China (NSFC) under Contracts Nos. 11635010, 11935015, 11935016, 11935018, 12025502, 12035009, 12035013, 12061131003, 12192260, 12192261, 12192262, 12192263, 12192264, 12192265, 12221005, 12225509, 12235017, 12361141819; the Chinese Academy of Sciences (CAS) Large-Scale Scientific Facility Program; CAS under Contract No. YSBR-101; 100 Talents Program of CAS; The Institute of Nuclear and Particle Physics (INPAC) and Shanghai Key Laboratory for Particle Physics and Cosmology; ERC under Contract No. 758462; German Research Foundation DFG under Contract No. FOR5327; Istituto Nazionale di Fisica Nucleare, Italy; Knut and Alice Wallenberg Foundation under Contracts Nos. 2021.0174, 2021.0299; Ministry of Development of Turkey under Contract No. DPT2006K-120470; National Research Foundation of Korea under Contract No. NRF-2022R1A2C1092335; National Science and Technology fund of Mongolia; Polish National Science Centre under Contract No. 2024/53/B/ST2/00975; STFC (United Kingdom); Swedish Research Council under Contract No. 2019.04595; U. S. Department of Energy under Contract No. DE-FG02-05ER41374

\bibliography{bibliography.bib}

\clearpage
\onecolumngrid
\appendix
\section*{Appendix}
\label{supplemental}
\renewcommand{\thefigure}{\alph{figure}} 
\setcounter{figure}{0} 

\begin{figure}[htbp]
\centering
\includegraphics[width=1.0\linewidth]{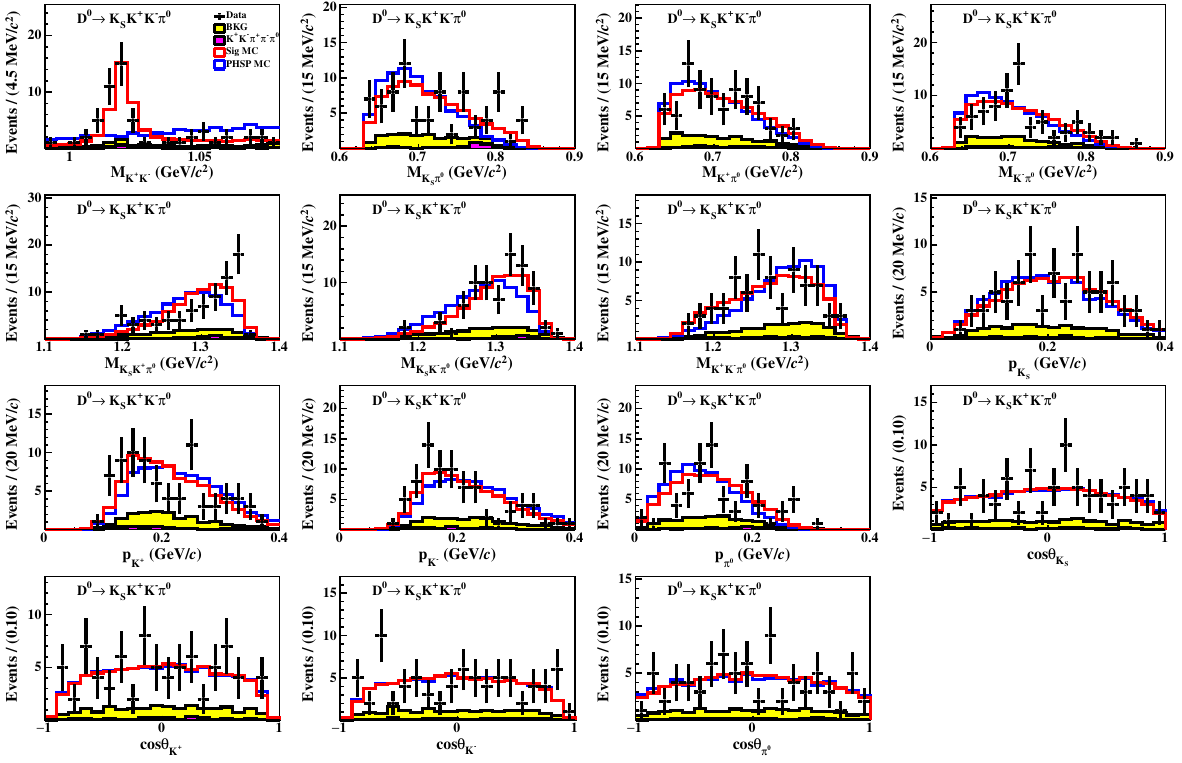}

\caption{
Comparisons of some typical distributions for the $D^0\to K_S^0K^+K^-\pi^0$ candidate events between data (dots with error bars) and 
and the signal MC sample (red and blue histograms), where the red one shows the mixed signal MC sample and the blue one the PHSP MC. The MC-simulated backgrounds~(BKG) from the inclusive MC sample are shown as yellow filled histograms.
}
\label{compare_6}
\end{figure}

\begin{figure}[htbp]
\centering
\includegraphics[width=1.0\linewidth]{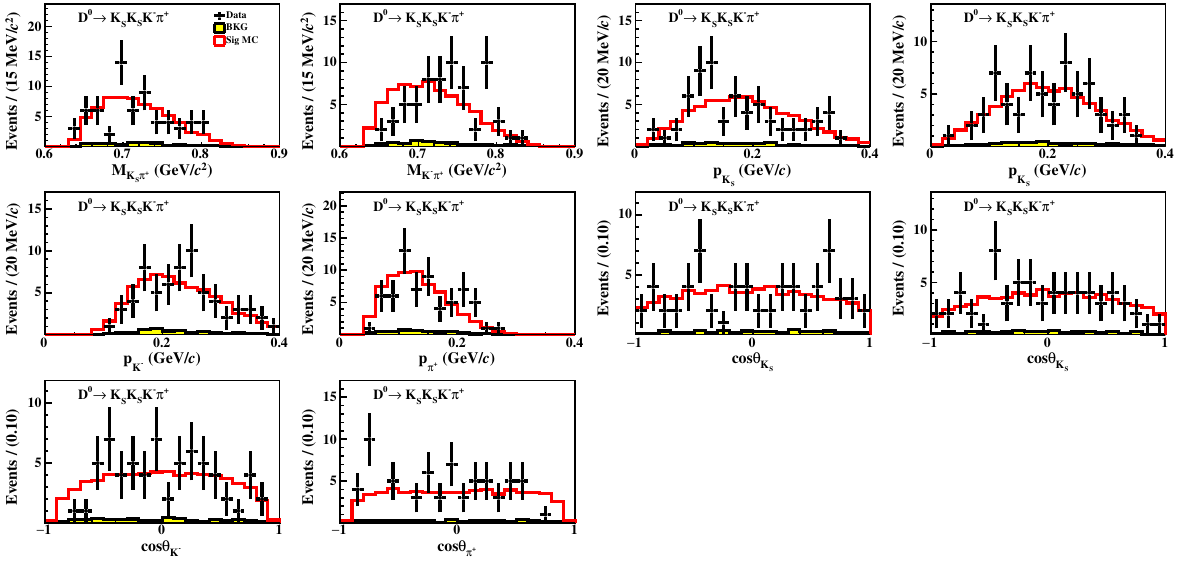}

\caption{
Comparisons of some typical distributions for the $D^0\to K_S^0K_S^0K^-\pi^+$ candidate events between data (dots with error bars) and the PHSP signal MC sample (red histograms). The MC-simulated backgrounds~(BKG) from the inclusive MC sample are shown as yellow filled histograms. 
}
\label{compare_7}
\end{figure}

\begin{figure}[htbp]
\centering
\includegraphics[width=1.0\linewidth]{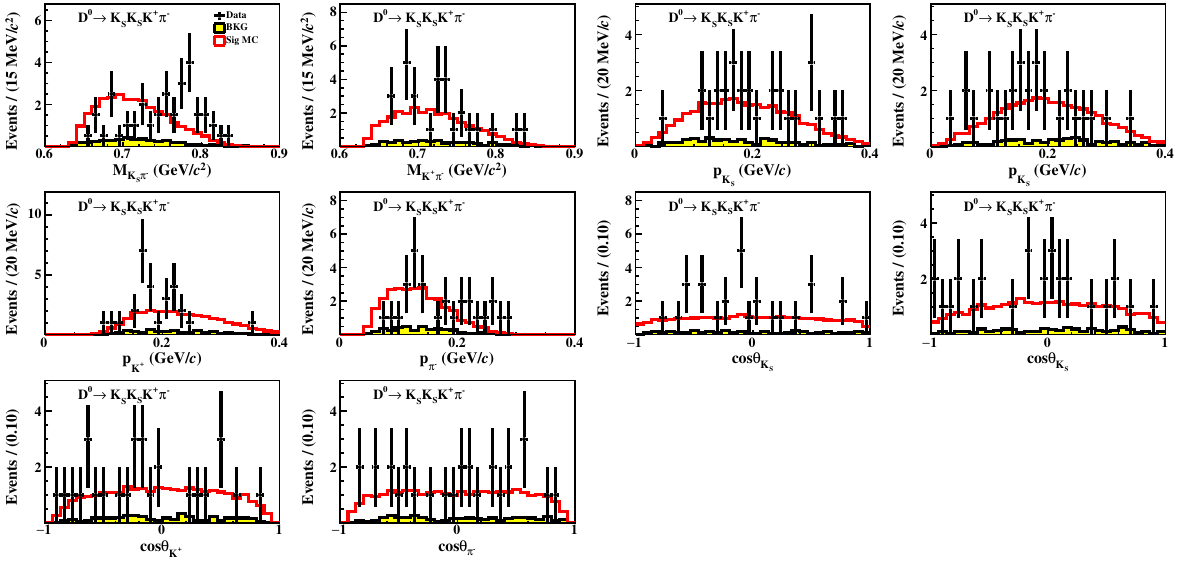}

\caption{
Comparisons of some typical distributions for the $D^0\to K_S^0K_S^0K^+\pi^-$ candidate events between data (dots with error bars) and the PHSP signal MC sample (red histograms). The MC-simulated backgrounds~(BKG) from the inclusive MC sample are shown as yellow filled histograms.
}
\label{compare_8}
\end{figure}

\begin{figure}[htbp]
\centering
\includegraphics[width=1.0\linewidth]{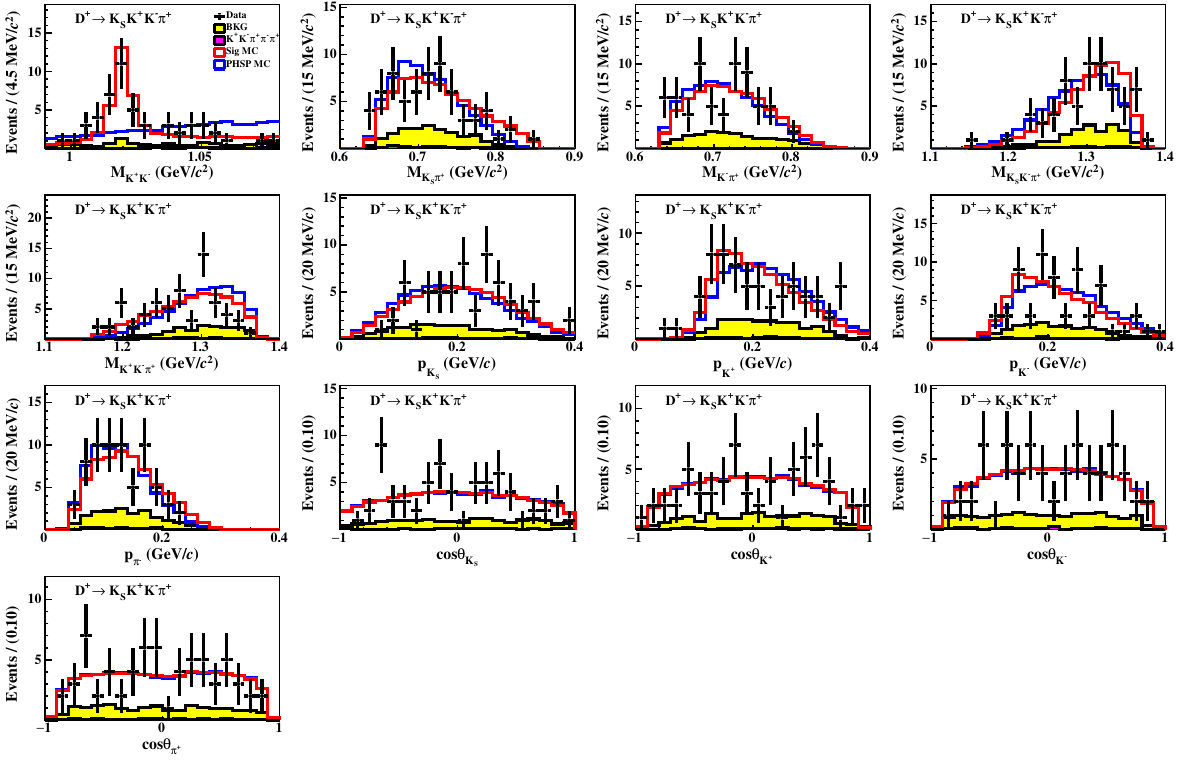}

\caption{
Comparisons of some typical distributions for the $D^+\to K_S^0 K^+K^-\pi^+$ candidate events between data (dots with error bars)
and the signal MC sample (red and blue histograms), where the red one shows the mixed signal MC sample and the blue one the PHSP MC. The MC-simulated backgrounds~(BKG) from the inclusive MC sample are shown as yellow filled histograms.
}
\label{compare_2}
\end{figure}

Figures~\ref{compare_6}, \ref{compare_7}, \ref{compare_8}, and \ref{compare_2} show the comparisons of the distributions of the momenta and cosines of  the polar angles of the daughter particles, as well as
the invariant masses of two-body or three-body particle combinations of the candidates for $D^0\to K_{S}^0K^+K^-\pi^0$, $D^0\to K_{S}^0K_{S}^0K^-\pi^+$, $D^0\to K_{S}^0K_{S}^0K^+\pi^-$, and $D^+\to K_{S}^0K^+K^-\pi^+$ between data and MC simulation, respectively.

\end{document}